\documentclass[prd,aps,superscriptaddress,twocolumn,nofootinbib]{revtex4}

\usepackage{graphicx}
\usepackage{amsmath,amssymb}
\usepackage{mathrsfs}
\usepackage[caption=false]{subfig}
\usepackage{hyperref}
\usepackage[x11names]{xcolor}
\usepackage{physics}
\usepackage[capitalize]{cleveref}

\def\iteq{\textit{e.g.}\,}
\def\itie{\textit{e.g.}\,}

\graphicspath{
    {figures/}
}

\begin{document}

\title{Repeated Gravitational Wave Bursts from Cosmic Strings}

\author{Pierre Auclair}
\email{pierre.auclair@uclouvain.be}
\affiliation{Cosmology, Universe and Relativity at Louvain (CURL),
Institute of Mathematics and Physics, University of Louvain, 2 Chemin du Cyclotron,
1348 Louvain-la-Neuve, Belgium}

\author{Dani\`ele A.~Steer}
\affiliation{Universit\'e Paris Cit\'e, CNRS, Astroparticule et Cosmologie, F-75013 Paris, France
}
\author{Tanmay Vachaspati}
\affiliation{
Physics Department, Arizona State University, Tempe,  Arizona 85287, USA}
\affiliation{
Universit\'e de Gen\`eve, D\'epartement de Physique Th\'eorique and Centre for Astroparticle
Physics, 24 quai Ernest-Ansermet, CH-1211 Gen\`eve 4, Switzerland
}

\begin{abstract}
A characteristic observational signature of cosmic strings are short-duration gravitational wave (GW) bursts. These have been searched for by the LIGO-Virgo-KAGRA (LVK) Collaboration, and will be searched for with LISA. We point out that these burst signals are repeated, since cosmic string loops evolve quasiperiodically in time, and will always appear from essentially the same position in the sky.
We estimate the number of GW repeaters for LVK and LISA, and show that the string tension that can be probed scales as detector sensitivity to the sixth power, which raises hope for detection in future GW detectors.
The observation of repeated GW bursts from the same cosmic string loop helps distinguish between the GW waveform parameters and the sky localization.

\end{abstract}

\maketitle

Symmetry breaking phase transitions in the early Universe may lead to the formation of topological defects of different kinds, see \iteq\cite{Vilenkin:2000jqa}. In this paper we focus on linelike defects, namely, cosmic strings \cite{Kibble:1976sj,Hindmarsh:1994re}, whose dimensionless  energy per unit length $G\mu$ (where $G$ is Newton's constant) is given by
\begin{equation}
	G\mu \sim 10^{-6} \qty(\frac{\eta}{10^{16} ~\mathrm{GeV}})^2
\end{equation}
where $\eta$ is the energy scale of the phase transition.  Cosmic strings may leave a variety of observational signatures (see \iteq\cite{Vilenkin:2000jqa,Vachaspati:2015cma,Ringeval:2010ca}), and here we focus on gravitational waves (GWs).  As is well known, closed loops of cosmic strings emit GWs and decay. The overall signature of these decaying loops (which are continuously created by the cosmic string network) is the generation of a stochastic GW background (SGWB), which spans many decades in frequency, see \iteq\cite{Auclair_2020}.
Current constraints on $G\mu$ from the SGWB depend on the loop distribution, and are $G\mu \lesssim 10^{-8}$ for the LIGO-Virgo-KAGRA (LVK) Collaboration~\cite{LIGOScientific:2021nrg} and $G\mu \lesssim 10^{-10}$ for Pulsar Timing Arrays ~\cite{Ellis:2020ena,Blasi:2020mfx,Bian:2022tju,Chen:2022azo,EPTA:2023hof}.

A second GW signal is in the form of bursts from loops which can be individually resolved~\cite{Damour:2000wa,Damour:2001bk}.
Indeed, closed loops of cosmic strings can contain cusps, namely points at which the string instantaneously moves at the speed of light. Furthermore, cusps are expected to be present on the vast majority of loops~\cite{Copi:2008he,Blanco-Pillado:2015ana}.
These cusps are well-known sources of short-duration, linearly polarized GW bursts, and have been searched for by the LVK Collaboration~\cite{LIGOScientific:2017ikf,LIGOScientific:2021nrg}.

In this paper, we point out another feature of these cusp signals that has not, to the best of our knowledge, been addressed so far. Indeed, a consequence of the Nambu-Goto equations in flat space is that a sub-horizon loop of invariant length $\ell$ evolves periodically in time with period $\ell/2$.  A GW burst from a cusp should therefore repeat in time, provided backreaction effects do not affect its dynamics significantly. We estimate the GW backreaction and show that it has a negligible effect on the beam direction. Hence, the observed beam will essentially always come from  the same position in the sky.
This repeating signal (as opposed to the continuous signal considered in Ref.~\cite{Dubath:2007wu}\footnote{In Ref.~\cite{Dubath:2007wu} ``periodic'' refers to a continuous oscillatory signal, coming from the lowest harmonic modes of a highly boosted loop. The cusp signal we consider is not continuous, and is a result of the higher harmonic modes \cite{Vilenkin:2000jqa}.}) can increase the detectability of the bursts and can enable better estimates of the cosmic string loop parameters (loop length, redshift and sky localization).
We calculate the typical rates and periods of the repeating cusps one could detect with LVK and LISA. We also show how these rates scale with the sensitivity of a (future) detector. A more detailed calculation together with corresponding data analysis will be presented elsewhere.

\section{GW emission from cosmic string cusps}
A cusp on a cosmic string loop of length $\ell$, located at a comoving distance $r(z) \gg \ell$, emits a short burst of gravitational radiation, whose waveform was calculated by Damour and Vilenkin~\cite{Damour:2000wa,Damour:2001bk} (see also Ref.~\cite{Siemens:2006vk}).
The Fourier transform
of the cusp waveform follows a power-law $h(f) = A f^{-4/3}$
with $f$ the frequency at the observer and
with an amplitude given by
\begin{equation}
    A = g_1 \frac{G\mu \ell^{2/3}}{(1+z)^{1/3} r(z)},
    \label{eq:Adef}
\end{equation}
where $g_1$ is a constant of order $1$, and $\ell$ is the invariant length
of the loop (defined by its total energy divided by $\mu$).
The detection and reconstruction of GW bursts by a GW interferometer is possible if the amplitude of the burst is larger
than $A_*$. This has been determined for LVK in Ref.~\cite{LIGOScientific:2021nrg} and for
LISA in Ref.~\cite{Auclair:2023brk}.
For LVK, with characteristic detector frequency $f_*\sim  20$ Hz, we have $A_* \sim 2\times 10^{-20} {\rm s}^{-1/3}$. For LISA, $f_*\sim 1$ mHz, $A_* \sim 3\times 10^{-21} {\rm s}^{-1/3}$, and the duration of typical bursts is of order an hour; see Fig.~2 in Ref.~\cite{Auclair:2023brk} or Fig.~1 in Ref.~\cite{ShapiroKey:2008ckh}.

The emission of GWs is concentrated in a beam of half-angle
\begin{equation}
    \theta_\mathrm{m} = [g_2 f (1+z) \ell]^{-1/3},
\end{equation}
where $g_2 = \sqrt{3}/4$ \cite{Auclair_2020}.
Note that $\theta_\mathrm{m}$ is larger at low frequencies and that the beam covers a larger
portion of the sky which enhances the probability that it is directed toward the Earth.

Loops smaller than the Hubble horizon can effectively be described with Nambu-Goto equations of motion in Minkowski space-time, in which case they oscillate with period $T = \ell/2$.
Therefore, one can expect to see multiple repetitions of GWs bursts from the same cosmic string loop.

As a rule of thumb, suppose that we are interested in a loop of length $\ell = 2$ ly so that its period is of one year.
Assuming the redshift $z < 1$ so that $r(z) \approx z / H_0$ with $H_0$ the Hubble constant; then it follows from Eq.~\eqref{eq:Adef} that a GW detector may detect cosmic string cusp bursts up to
\begin{equation}
    z_*(\ell) = \frac{g_1 G\mu \ell^{2/3} H_0}{A_*} \approx 0.35 ~ \left(\frac{\ell}{2 \mathrm{yr}}\right)^{2/3}\frac{G\mu}{10^{-9}} \frac{10^{-21}}{A_*}.
    \label{eq:zs}
\end{equation}
For the most remote detectable loops, with $z=z_* < 1$, the beaming angle is of the order of
\begin{equation}
    \theta_\mathrm{m} \approx 3\times 10^{-3} \left(\frac{1 \mathrm{Hz}}{f_*} \times \frac{2 \mathrm{yr}}{\ell}\right)^{-1/3}
    \label{eq:theta}
\end{equation}
where $f_*$ is the characteristic detector frequency, namely, $f_*\sim  20$ Hz for LVK and $f_*\sim 1$ mHz for LISA.

\section{Repeated bursts}
As we mentioned earlier, a Nambu-Goto loop of invariant length $\ell$ oscillates with period $T=\ell/2$ and cusps reappear every oscillation.
However, physical loops are not exactly periodic as they lose energy in the form of GWs and particles, particularly around the cusp.
As a result, after one oscillation, the beaming direction will be slightly displaced due to the backreaction of the beam on the dynamics of the loop.
We now estimate the change in the beaming direction, $\Delta\theta$, after a GW burst from a cusp.

Field theory simulations of cosmic strings~\cite{Blanco-Pillado:1998tyu,Olum:1998ag} show that the string at the very tip of the cusp annihilates during the burst, leading to the emission of particles, which subsequently decay into cosmic rays and photons. However, the portion of the string responsible for gravitational emission is farther away from the tip of the cusp, which follows the Nambu-Goto trajectory very closely, and as a consequence the GW burst repeats with similar amplitude.
Our aim in this section is to determine whether the  angular momentum carried away by the GW burst will change the orientation of the cusp: if it does, a repeating GW burst might not be visible.

The angular momentum of the loop can be estimated as a product of its moment of inertia,
$I_\ell \sim \mu \ell^3$, and the angular frequency $\omega \sim 1/\ell$,
\begin{equation}
    J \sim I_\ell \omega   \sim \mu \ell^2 .
\end{equation}
The cusp GW burst carries with it some of this angular momentum~\cite{Durrer:1989zi} \footnote{Some angular momentum would be lost due to cusp annihilation, but this is suppressed by a factor $\ell^{-1/2} \mu^{-1/4} $ \cite{Blanco-Pillado:1998tyu}.}. The energy of the burst in GWs
is $\sim G\mu^2 \ell $ and so the angular momentum emitted in a burst is
\begin{equation}
    \Delta J \sim G\mu^2 \ell ^2,
\end{equation}
and the rotation of the cusp beam direction in one time period follows from
\begin{equation}
    I_\ell  \dot\theta \sim \Delta J
\end{equation}
or
\begin{equation}
    \Delta \theta \sim (G\mu^2 \ell ^2) \, \frac{\ell}  { (\mu \ell ^3)} \sim G\mu .
\end{equation}
Since we are interested in $G\mu \lesssim 10^{-9}$, we see from \cref{eq:theta} that
\begin{equation}
    \Delta \theta \ll \theta_m(f_*,z)
\end{equation}
for all $f_*$, $z$, and $T$ of interest,
and the change in the beaming direction is negligible.
This is the reason why gravitational wave bursts from cosmic string cusps will appear as repeaters.

\begin{figure}
    \includegraphics[width=.45\textwidth]{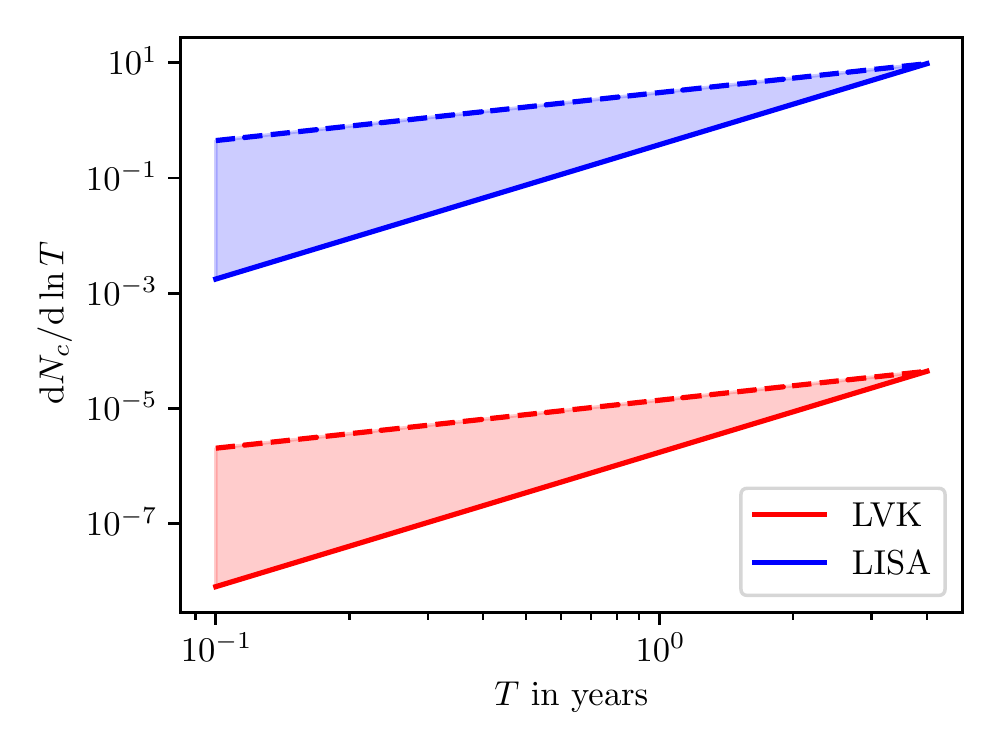}
    \caption{
    Number of repeaters per logarithmic bin of the period, $T=\ell/2$, as seen by different GW detectors for $G\mu=10^{-10}$.
    The solid line shows \cref{eq:rate-nogain}, ignoring repetitions. The dashed line shows \cref{eq:rate-gain}, namely including the sensitivity gain due to repetition.
    If not all repeaters are observed, one would expect a rate somewhere in the shaded regions.
    (Assuming $T_O = 4$ years for each detector.)}
    \label{dNdlogTvsT}
\end{figure}

\begin{table}
    \centering
    \begin{ruledtabular}
        \begin{tabular}{ccc}
            & \textbf{LVK O3} & \textbf{LISA}
            \\
            $A_*$ & $2 \times 10^{-20} \mathrm{s}^{-1/3}$ & $3 \times 10^{-21} \mathrm{s}^{-1/3}$
            \\
            $f_*$ & $20$ Hz & $1$ mHz
            \\
            \hline
            \textbf{Single event} \\
            ${\dv*{N_c}{\log T}} (1 \mathrm{mo})$ & $5 \times 10^{-9}$ & $1 \times 10^{-3}$
            \\
            ${\dv*{N_c}{\log T}} (1 \mathrm{yr})$ & $2\times 10^{-6}$ & $0.4$
            \\
            \hline
            \textbf{With repetitions} \\
            ${\dv*{N_c}{\log T}} (1 \mathrm{mo})$ & $2 \times 10^{-6}$ & $0.4$
            \\
            ${\dv*{N_c}{\log T}} (1 \mathrm{yr})$ & $1 \times 10^{-5}$ & $3$
        \end{tabular}
    \end{ruledtabular}
    \caption{Characteristics of current and future GW detectors and predicted number of repeaters with period of one month and one year for $G\mu = 10^{-10}$.
    }
    \label{tab:tab}
\end{table}

\section{Number of repeaters}
To compute the number of repeaters, let us assume that the distribution of cosmic string loops in the matter era is given by the one-scale model \cite{Vilenkin:2000jqa},
\begin{equation}
	{\pdv[2]{N}{\ell}{V}}(\ell,t)
	= \frac{C_\mathrm{mat}}{t^2 (\ell+\Gamma G\mu t)^2}
	+ \frac{C_\mathrm{rad}}{t_\mathrm{eq}^{3/2} (\ell+\Gamma G\mu t)^{5/2}} \left ( \frac{t_\mathrm{eq}}{t} \right )^2,
\label{dn}
\end{equation}
where the first term is the contribution from loops produced during the matter era and the second
term is the contribution from loops surviving from the radiation era into the matter era.
$\Gamma = \order{50}$ is a numerical factor that quantifies the efficiency of GW emission and $C_\mathrm{rad} = 0.18$~\cite{Blanco-Pillado:2013qja}.
In our estimate, we assume that the second contribution dominates, as is commonly assumed and that loops have on average two cusps~\cite{Blanco-Pillado:2015ana}.

Since the GW emission is beamed into a cone of half-angle $\theta_\mathrm{m}$, the number of cusps directed toward the Earth is
\begin{equation}
    \pdv[2]{N_c}{\ell}{V}(\ell, t) = \frac{2 \pi \theta_\mathrm{m}^2}{4\pi} {\pdv[2]{N}{\ell}{V}}
    \approx \frac{\theta_\mathrm{m}^2}{2}\frac{C_\mathrm{rad}}{t_\mathrm{eq}^{3/2} (\ell+\Gamma G\mu t)^{5/2}} \left ( \frac{t_\mathrm{eq}}{t} \right )^2.
\end{equation}
Only those cusps that are located closer than $z_*(\ell)$ can be detected in our detector.
Since we showed in \cref{eq:zs} that $z_*$ is generally smaller than $1$, we assume in the following $z_* \ll 1$ and set $t = t_0 = 3 \times 10^{17}$s, the age of the Universe.
Taking as a volume element
\begin{equation}
    \dd{V} = \frac{4 \pi r^2(z) \dd{z}}{(1+z)^3 H(z)} \underset{z < 1}{\approx} \frac{4 \pi z^2 \dd{z}}{H_0^3},
    \label{eq:dv}
\end{equation}
the number of detectable cusps with period $\ell/2$ is given by
\begin{align}
    {\dv{N_c}{\ell}} (\ell) =& \int_0^{z_*} \pdv[2]{N_c}{\ell}{V} \dd{V} \nonumber \\
    \underset{z_*<1}{\approx}& \frac{2 g_1^3\pi}{3 g_2^{2/3}} \frac{C_\mathrm{rad} t_0^{-3/2} \ell^{4/3}}{(\ell+\Gamma G\mu t_0)^{5/2}} \sqrt{\frac{t_\mathrm{eq}}{t_0}} \frac{(G\mu)^3}{A_*^3 f_*^{2/3}}.
\end{align}
Each of the $N_c$ cusps will be detected as GW repeaters.
Notice that the number of repeaters depends strongly on the detection threshold $A_*$ and only weakly on the detector's frequency.
For $\ell = \order{\mathrm{yr}}$ and $\Gamma G\mu > 10^{-10}$, then $\Gamma G\mu t_0 \gg \ell$ and the number of repeaters can be simplified to
\begin{equation}
    {\dv{N_c}{\ell}} (\ell) \underset{z_*<1}{\approx} \frac{2 g_1^3\pi C_\mathrm{rad} t_0^{-4} \ell^{4/3}}{3 g_2^{2/3} \Gamma^{5/2}} \sqrt{\frac{t_\mathrm{eq}}{t_0}} \frac{\sqrt{G\mu}}{A_*^3 f_*^{2/3}},
    \label{eq:rate-nogain}
\end{equation}
and the number of repeaters is not very sensitive to the string tension $G\mu$.
In \cref{dNdlogTvsT}, we show the number of repeaters per logarithmic bin of period for both LISA and LVK, and \cref{tab:tab} contains the numerical values for the number of repeaters with period one month and one year.

As an order of magnitude, the number of repeaters per logarithmic unit of $\ell$ for $\ell = 2$ yr,
\itie, time period of one year, normalized with the parameters of LISA is
\begin{equation}
    \eval{\dv{N_c}{\log T}}_\mathrm{1 yr}
    \underset{z_*<1}{\approx} 4~ \sqrt{\frac{G\mu}{10^{-9}}}\left(\frac{2\times 10^{-21}}{A_*}\right)^3 \left(\frac{1\, \mathrm{mHz}}{f}\right)^{2/3}.
\end{equation}
This estimate gives the number of sources that will be repeating with one year period.

\begin{figure*}
	\subfloat[]{\includegraphics[width=.45\textwidth]{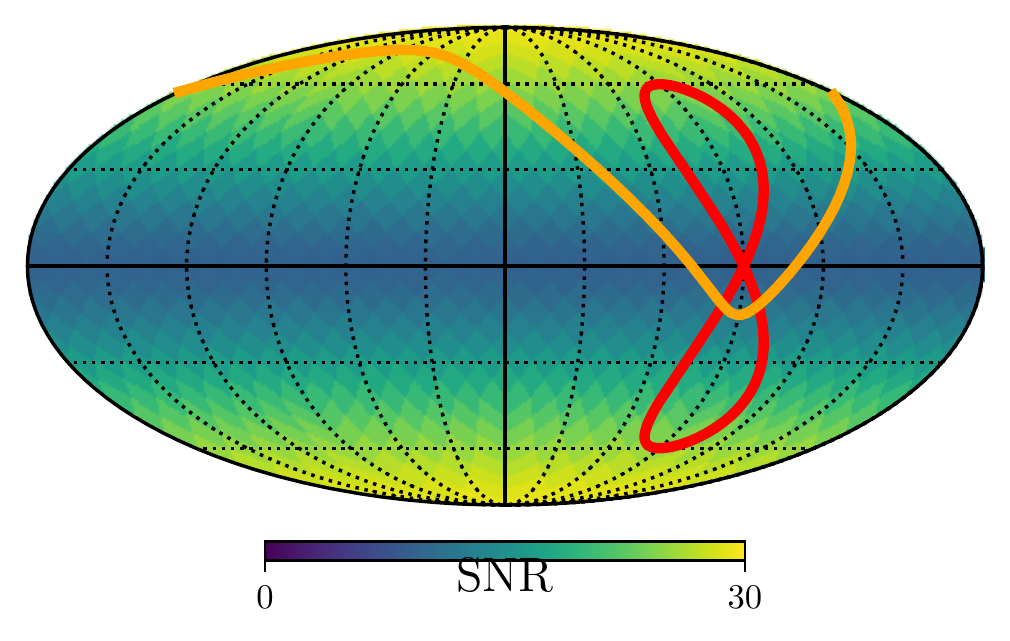}}
    \subfloat[]{\includegraphics[width=.45\textwidth]{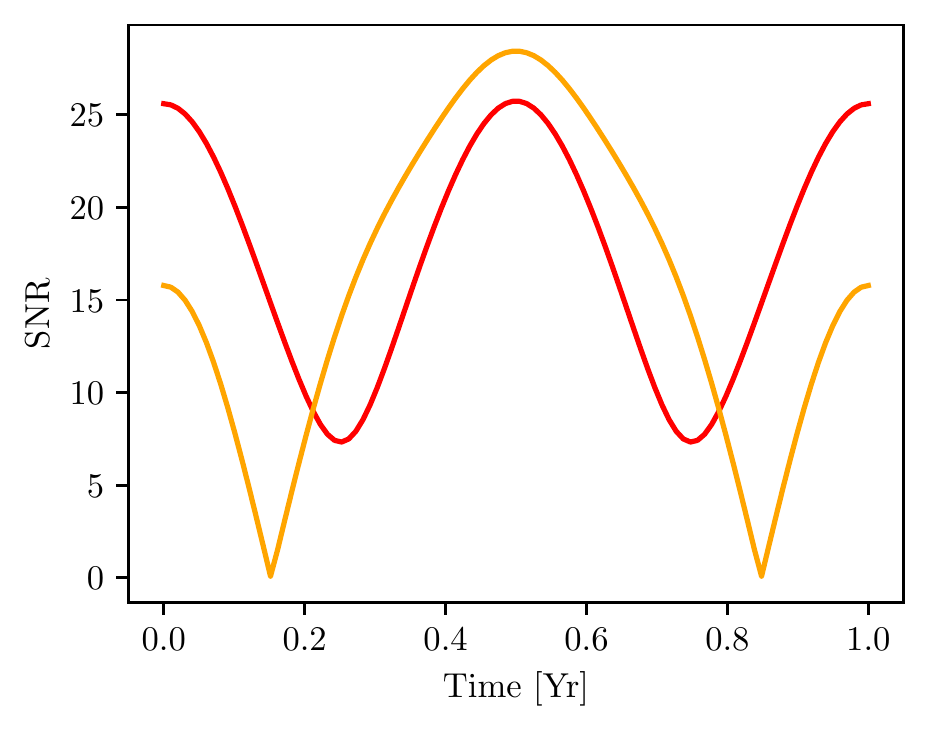}}
    \caption{Modulation of the SNR in the reference frame of LISA for a repeating burst of amplitude $A = 10^{-21}$ in the equatorial plane (red) and at colatitude $\pi / 4$ (orange) and with same longitude. Left panel: antenna pattern of LISA in its own reference frame using Time Delay Interferometry (TDI) and generation 1.5 variables~ \cite{Tinto:2020fcc,Auclair:2023brk}. Right panel: evolution of the SNR over the course of one year.}
    \label{fig:modulation}
\end{figure*}

\section{Detection}

A key point is that the combined waveform of the collection of bursts will be periodic with period $T = \ell / 2$ and this periodicity may help extract (and confirm) the signal.  Recall that with LISA parameters the duration of a burst is of order an hour \cite{ShapiroKey:2008ckh,Auclair:2023brk}, whereas the repeating time is of order months to years. The bursts therefore do not overlap.

One approach to the data analysis for gravitational wave repeaters is similar to that for
fast radio burst (FRB) repeaters, \iteq, Ref.~\cite{Morello_2020}).
Note that in the context of FRBs the burst lasts for several milliseconds, and there are of order several FRBs detected per day,
whereas in context of GW from cosmic strings in LISA, the bursts last for several minutes to hours~\cite{ShapiroKey:2008ckh}, and there are of order several bursts per year.
In both cases, there is a separation of scales between the duration of the bursts and their occurrence, hence suppressing the probability of overlap.
To search for a repeating
signal with period $T$ in a time series data, one can divide the data in $N$ bins each
of temporal length $T$, and then average the data over the $N$ bins. If there is a signal
with period $T$, it will add up coherently, giving an average amplitude $S$ where $S$
is the signal of a single event. The noise in the data will add up incoherently and be
suppressed by a factor of $1/\sqrt{N}$.

This gives a gain in sensitivity by the factor $\sqrt{N}$, where $N$ is the number of
repetitions that are observed, \itie, $N = \max(1, T_O/T)$ where $T_O$ is the observation
time.
Since we are interested in repeating bursts, we restrict ourselves to $N=T_O / T > 1$.

As a proxy, we can then define an ``effective'' and period-dependent amplitude (denoted by a tilde) of the repeated burst as
\begin{equation}
\tilde{A}_* = \frac{A_*}{\sqrt{N}} = A_* \sqrt{\frac{\ell}{2 T_0}}.
\label{tildeA}
\end{equation}
Consequently, the number of detectable repeaters \emph{assuming} this gain of sensitivity of $\sqrt{N}$ (denoted by a tilde again) is modified to
\begin{equation}
    {\dv{N_c}{\ell}} (\ell) \underset{z_*<1}{\approx} \frac{2^{5/2} g_1^3\pi C_\mathrm{rad} t_0^{-4} T_0^{3/2}}{3 g_2^{2/3} \Gamma^{5/2} \ell^{1/6}} \sqrt{\frac{t_\mathrm{eq}}{t_0}} \frac{\sqrt{G\mu}}{A_*^3 f_*^{2/3}}.
    \label{eq:rate-gain}
\end{equation}
This last equation is shown as dashed lines for both LISA and LVK in \cref{dNdlogTvsT}, and the numerical values for one month and one year are presented in \cref{tab:tab}.
While this rate remains similar for repeaters with period the order of $T_O$, this gain of sensitivity boosts repeaters with smaller periods.
Note that the dependence on the string tension $G\mu$ is degenerate with $A_*^6$; hence mild improvements on the detector's sensitivity have an important impact on $G\mu$.

A second key point for us is that the repeated observation of the same burst over the course of a year (for LISA) or of days (for ground-based detectors) may help lift the degeneracy between the amplitude of the waveform $A$ and the sky localization of the source and help with parameter estimation (see Refs.~\cite{Babak:2008aa,Cohen:2010xd,ShapiroKey:2008ckh, MockLISADataChallengeTaskForce:2009wir} for cosmic string burst parameter estimation in the context of the Mock LISA Data Challenge 3).
Indeed, the signal-to-noise ratio (SNR) of the repeated bursts is modulated by the response function of the detector.
As an example, \cref{fig:modulation} shows the SNR modulation of the same GW bursts over the course of a year in LISA if the source is on the plane of the ecliptic or at colatitude $\pi/4$.

Finally, Collaboration with other space-borne GW detectors such as TianQin~\cite{TianQin:2020hid} could be greatly helpful in identifying the astronomical nature of a burst signal.

\section{Conclusion}
Our plot in Fig.~\ref{dNdlogTvsT} shows that LISA will be sensitive to repeated bursts from
cosmic string cusps for $G\mu =10^{-10}$, even for bursts that repeat on a monthly period
or larger.
Our estimate assumes two cusps on average per loop. If the number of cusps per loop were to increase the number of repeaters would increase proportionally.
As \eqref{eq:rate-gain} shows, the number of repeaters scales as $\sqrt{G\mu}$ and is most sensitive
to the detection
threshold since it goes as $A_*^{-3}$.
A modest improvement in detector sensitivity would result in constraints on $G\mu$ that scale as $A_*^6$.
The dependence on the detector frequency is
milder, going as $f_*^{-2/3}$.

Gravitational wave emission from a cusp may also be accompanied by the emission of photons \cite{Steer:2010jk}, and of
cosmic rays due to annihilation of the very tip of the cusp~\cite{Blanco-Pillado:1998tyu,Olum:1998ag}. Whether the cosmic
ray emission is also periodic is not known at the moment because it depends on the backreaction of cusp radiation on the dynamics of the string. If a smooth loop develops a cusp, the tip of the cusp annihilates and the string
obtains two kinks. If radiation backreaction is not effective in smoothing
out the kinks, in the next period the Nambu-Goto evolution of the string will not lead
to reformation of the tip of the cusp. Then there will be no tip to annihilate, and cosmic-ray emission will not repeat. This argument does not apply to gravitational wave
emission because these originate farther away from the cusp and not from the tip.
In certain other cases, for example if a scalar field condenses on the string or if the
string is superconducting and has weak currents, there might be repeated nongravitational
emission in the form of scalar particles~\cite{Vachaspati:2009kq} and radio bursts~\cite{Vachaspati:2008su,Cai:2012zd}.
These considerations open up
the possibility of multimessenger signatures for cosmic strings, but the details need further work.

Cosmic strings have other sharp features called kinks that also radiate gravitational waves, less intensely than cusps but the number of repeaters may be higher as there are more kinks, and they emit in larger beams~\cite{Garfinkle:1987yw}. It would be worth investigating repeated bursts from kinks and kink-kink collisions.

There are several complications that will occur in a realistic setting.
First is that the detector moves over time and different occurrences of the repeater will not provide identical signals; therefore the identification of the repeated signals will require new data analysis techniques.
Second is that GW detectors typically do not take data continuously but rather have a sparse data stream due to maintenance and improvement breaks.
For instance, LISA is expected to have a duty cycle of about $82\%$.
Therefore, some repetitions of the GW bursts may be outside the covered time intervals.

\begin{acknowledgments}
We thank Lars Aalsma, Stas Babak,
Jose Juan Blanco-Pillado, Simon Foreman, Tjonnie Li, Ken Olum, and Gage Siebert for comments. We also thank Chiara Caprini and Ruth Durrer for hospitality at
the University of Geneva. We gratefully acknowledge CERN for hospitality
while this work was in progress, and
the ``Superconducting strings'' workshop at IAP in
December 2022 that was a starting point for this project.
This work was supported by the U.S. Department of
Energy, Office of High Energy
Physics, under Award No.~DE-SC0019470.
The work of P.A. is supported by the Wallonia-Brussels
Federation Grant ARC \textnumero~19/24-103.
\end{acknowledgments}

\bibstyle{aps}
\bibliography{tvNotes}

\end{document}